\documentclass[journal]{IEEEtran}
\pdfoutput=1
\usepackage[pdftex]{graphicx}
\usepackage{amsmath}

\usepackage{eqparbox}
\usepackage{float}
\usepackage{stfloats}
\usepackage{booktabs}
\usepackage{setspace}
\usepackage{amssymb}
\usepackage{caption}
\usepackage{algorithm}
\usepackage{algpseudocode}
\usepackage{multirow}{\centering}
\usepackage{subfig}
\usepackage{subfloat}
\usepackage{url}
\usepackage{array}
\usepackage{diagbox}
\usepackage{makecell}
\usepackage{color}
\usepackage{amsmath}
\usepackage{bm}
\begin{document}
\captionsetup[figure]{labelfont={bf},name={Fig.},labelsep=period}
\captionsetup[table]{labelfont={bf},labelsep=period}

\title{Deep Open Set Identification for RF Devices \\
\thanks{This work is financially supported by the National Natural Science Foundation of China under Grant No. 61871282. (\textit{Corresponding author: Haoyu Fang.})}
\thanks{Qing Wang, Qing Liu and Zihao Zhang are with the school of Electrical and Information Engineering, Tianjin University, Tianjin 300072, China(email: wangq@tju.edu.cn,
liuqinglq@tju.edu.cn, zhangzihao@tju.edu.cn).}
\thanks{Haoyu Fang is with the Electrical and Computer Engineering, New York University, New York, USA(email: haoyu.fang@nyu.edu).}
\thanks{Xi Zheng is with the Department of Computing, Macquarie University, NSW, Australia(email:  james.zheng@mq.edu.au).
}}
\author{Qing~Wang,~\IEEEmembership{Member,~IEEE,}
Qing Liu, Zihao Zhang,\\ Haoyu Fang,~\IEEEmembership{Student Member,~IEEE},
Xi Zheng,~\IEEEmembership{Member,~IEEE}
}


\maketitle


\begin{abstract}
Artificial intelligence (AI) based device identification improves the security of the internet of things (IoT), and accelerates the authentication process.  
However, existing approaches rely on the assumption that we can learn all the classes from the training set, namely, closed-set classification.
To overcome the closed-set limitation, we propose a novel open set RF device identification method
to classify unseen classes in the testing set. First, we design a specific convolution neural network (CNN) with a short-time Fourier transforming (STFT) pre-processing module, which efficiently recognizes the differences of feature maps learned from various RF device signals. Then to generate a representation of known class bounds, we model the peripheral samples' distribution and revise CNN's output according to the extreme value theory (EVT). Finally, we estimate the probability map of the open-set via the OpenMax function.
We conduct experiments on sampled data and voice signal sets, considering various pre-processing schemes, network structures, distance metrics, tail sizes, and openness degrees. The simulation results show the superiority of the proposed method in terms of robustness and accuracy.
\end{abstract}



\section{Introduction}

With the development of massive accesses to the internet of things (IoT), radio frequency (RF) device identification provides critical security ability \cite{mscnn}.
RF device identification can recognize devices by radio-frequency signals instead of password-based security protocol recognizing the primary user’s signal, preventing the primary user emulation attack (PUEA) \cite{PUEA}. Besides, RF device identification protects communication networks from unauthorized access, association with rogue access points (APs), masquerading attack, and a host of other invasions \cite{na3}. Therefore, it is important to improve the performance of RF individuals' identification in the communication network.

The outstanding classification performance of deep learning \cite{deeplearning} has been applied to RF device identification and made great achievements \cite{DeepRFID}. However, during the past decades, researchers rely on the assumption that all classes have known as a prior, namely the closed set, which is not realistic. For the closed set classification problem, classes in the testing set are the same as in the training set. But for the RF device classification problem, the limited training set can not emulate all the RF device classes, resulting in incomplete prior information in the training phase. Thus, in the testing phase, the potential unknown class data can not be classified into any trained classes. Consequently, a new algorithm for open set recognition that can classify known RF devices and identify unknown classes is in demand. Therefore, we propose to design a deep open set classifier and use it to solve the open set RF device identification problem.

In general, the traditional deep neural network is incapable of identifying open sets with unknown classes. The last layer of deep neural networks contains a fully connected layer and a softmax function \cite{cnn1}, where the fully connected layer extracts features and the softmax function calculates the probability distribution of the features.
In the training phase, the network provides an n-dimensional probability distribution of the known classes. If the testing set contains untrained or unknown classes, the softmax function will estimate it as a certain known class with high confidence or all classes with low probabilities. In other words, the characteristic differences learned by a traditional deep neural network can be used for distinguishing known classes only.

The basic idea behind open set recognition is to allow rejection of artifacts
``fooling" 
to estimate the probability of an input being from an unknown class \cite{1vsset}. However, for the open set RF device identification problem, the differences between RF devices are slight. This leads to the first question addressed in this paper, ``How to adapt an open set classifier to the problem of open set RF device identification with slight differences?"

In this paper, by exploiting the ability of multi-scale features extraction and slight differences features expression, we first present a convolution neural network (CNN) with short-time Fourier transforming (STFT) pre-processing, yielding the best performance in close-set classification.
Open set recognition, i.e., identifying unknown classes, can be achieved through measuring the distances of feature values between the training samples and the corrected classified samples.
Moreover, we introduce the extreme value theory (EVT) to manage the peripheral feature values by modeling a Weibull distribution of the tail values; thus, we have a better expression of known class bounds than the hard thresholds.
Later, the revised weighted cumulative distribution function (CDF) of the Weibull distribution is used to calculate the open set feature values of testing samples, which contains a new $0$-th dimension to indicate the unknown class feature values.
In contrast to estimating the probability distribution belonging to known classes using the softmax function, we exploit unknown class features to estimate the open set probability using the openmax function. In addition, we study the performance of the proposed deep open set RF device identification method in different datasets, neural network models, distance metrics, the peripheral values tail sizes, and openness degrees.

Our main contributions include:
\begin{itemize}
\item We propose a novel open-set RF device identification framework that uses meta-learning theory to improve security.

\item We design a short-time Fourier transforming (STFT) pre-processing module to capture the slight differences among RF devices, integrating multi-scale features (e.g., time and frequency domain). Through experiments, we show the effectiveness of our pre-processing module.

\item To present the known class bounds more tightly, we calculate the Weibull distribution using extreme value theory. Then we calculate the open-set probability, namely open-set activation vectors (AV), to recognize the unseen classes.
\end{itemize}

The rest of the paper is organized as follows: Section \ref{relatework} introduces related works of device identification and open-set recognition. The open-set RF device identification method is proposed in Section \ref{method}. Section \ref{experiment} presents the preliminary and the experiment results, followed by the conclusion in Section \ref{conclusion}. And the definition of some important notations used in this paper are given in the following Tab \ref{notation}.
 \linespread{1.2}
    \begin{table}[h]
    \caption{Some important notations.
    \label{notation}}
    \centering
    \begin{tabular}{cc}
    \hline
    \textbf{Notations} & \textbf{Definition} \\
    \hline
    $\bm{s}$ & the training set  \\
    $\bm{s}^{\prime}$ &  the testing set \\
    $\bf{V}$ & the features of data \\
    $\bf{AVs}$ & the activation vectors \\
    ${\bf{v}}_{i,j}$ &      the $j$-th AVs of $i$-th class \\
    $\mathbb{V}$ &      \makecell[c]{the AVs of correctly\\ classified training samples} \\
    ${\bm{\mu}}_i$ & the mean activation vectors(MAV)\\
    $d_{i,j}$ & \makecell[c]{the distance between\\ the AVs and the corresponding MAV}\\
    $\omega$ & \makecell[c]{the Weibull probability\\ density function (PDF)}\\
    $k$ &  the shape parameter\\
    $\lambda$ & the scale parameter\\
    $\tau$ &  the location
    parameter\\
    $\Omega$ & \makecell[c]{the Weibull  cumulative\\ distribution function (CDF)}\\
    ${\Omega}^{\prime}_i$ & the revised weighted CDF\\
    $\eta_i$ &  the weighting parameter\\
    $\hat{\bf{v}}$ & the open-set AV\\
    $N_{TA}$ & the number of training classes\\ $N_{TE}$ & the numbers of the testing classes\\
    $N_{TG}$ & the numbers of the target classes\\
    $KP$ & the correctly classified known samples\\
    $KN$ & the incorrectly classified known samples\\
    $UP$ & the correctly classified unseen samples\\
    $UN$ & the incorrectly classified unseen samples\\
    \hline
    \end{tabular}
    \end{table}

\section{Related Works}\label{relatework}
\subsection{Device Identification}
Early device identification methods mainly rely on hand-extracted features. S. Jana extracted clock skew \cite{clockskew} from the timestamps of messages in medium access control (MAC) layer frames. A. K. Dalai used network and upper layer features like inter-arrival delay of network packets \cite{inter-arrival}. However, these features are easily forged, promoting us to extract physical layer (PHY) features. PHY features generally fall into two categories: location-dependent features and location-independent features. The location-dependent features are related to transceivers' environment or mobility, such as received signal strength (RSS) \cite{rss}. The location-independent features are inherent features related to hardware circuits, including transient and stationary features. As for the transient portion's duration is usually at nanosecond or sub-microsecond level, it is hard to be forged and can be used as a device fingerprint. Still, we need precise receivers to capture the transient features \cite{transientfeatures1} of amplitude, phase angle, and frequency \cite{transientfeatures2}. L. Peng designed a hybrid fingerprint extraction scheme to capture carrier frequency offset, I/Q offset, and other features \cite{stationaryfeatures}. G. Huang extracted amplifier nonlinearity characteristics \cite{amplifier1} caused by power amplifier imperfections \cite{amplifier2}. Some researchers use multiple features, which are cultivated from received signals \cite{multifeature} or robust principle component analyzed (RPCA) features \cite{pcafeature} to improve accuracy. However, it must be pointed out that the performances of these hand-extracted features based methods are limited. Existing fingerprint-based methods face the problem of the unstable region of interest (ROI), high-cost feature design, and incomplete automation \cite{mscnn}.

To address these problems, deep learning-based methods are proposed for feature exaction and classification automatically.
Authors in \cite{DeepRFID} use time-domain complex baseband error signals to train a convolutional neural network for cognitive radio device identification. Authors in \cite{deeprnn} proposed a long short-term memory based recurrent neural network for automatically identifying hardware-specific features and classifying transmitters in the presence of intense noise.
The multi-sampling convolutional neural network (MSCNN) is proposed to extract fingerprints using multiple downsampling transformations for multi-scale feature extraction \cite{mscnn}. Using the end-to-end network model, deep learning methods can overcome the limitation of hand-extracting features and achieve high recognition accuracy. However, these works are all under close-set condition, and cannot identify unseen devices.

\subsection{Closed-Set Recognition}
For the closed-set classification problem, classes in the testing set are the same as in the training set. Nevertheless, the limited training set can not emulate all the RF device classes for the RF device identification problem, resulting in incomplete prior information in the training process. Thus, the data of potential unseen classes can not be classified into trained classes in the testing process. Consequently, a new algorithm for open-set recognition, which can not only classify known RF devices but also identify unseen classes, is in demand. Therefore, we design a deep open-set classifier to solve the open-set RF device identification problem.

In general, the traditional deep neural network is incapable of identifying open-sets with unseen classes. The last layer of deep neural network contains a fully connected layer and a Softmax function \cite{cnn1}, where the fully connected layer extracts features and the Softmax function calculates the probability distribution of the features.

In the training process, the network provides an $n$-dimensional probability distribution of the known classes. If the testing set contains untrained or unseen classes, the Softmax function will estimate it as a particular known class with high confidence or all classes with low probabilities. In other words, the characteristic differences learned by a traditional deep neural network can be used for distinguishing known classes only.

\subsection{Open-Set Recognition}
	Open-set recognition deals with the problem that the finite training set can not emulate all kinds of possible classes \cite{opensetproblem1,opensetproblem2}. The term ``open-set" refers to the testing set containing data that doesn't belong to any trained classes.

    One of the open-set recognition methods introduces a threshold to the closed-set classifier \cite{oldrfid2,oldrfid3,evm}. The ``1-vs-Set Machine" method \cite{1vsset} adds a second hyper-plane to create a ``half-space" as an open space around the decision boundary for both 1-class and binary support vector machines (SVMs). By generalizing or specializing the two planes, open-set identification is changed to a risk-minimizing constrained functional optimization problem, but this method identify only one unseen class. Later, Walter J. \cite{wsvm} introduced non-linear classifiers Weibull-calibrated SVM (W-SVM) for multi-class open-set identification. They reduces a compact abating probability model (CAP), which reduce open space risk. $P_{I}$-SVM \cite{psvm} brings score calibration such as statistical EVT \cite{evt,evt2,evt3} to SVM. However, these SVM-based open-set classifiers have poor accuracy and high computational complexity.
	
	The deep learning method adopts the end-to-end model and has great performance in feature extraction. By setting a threshold on deep neural networks' output probabilities \cite{oldrfid1}, the sample is supposed to be unseen if all probabilities are below the threshold. However, a sample of an unseen class may have low probabilities of all classes. It's more like an uncertain class than an unseen class. Besides, the classifier network could be easily tricked by ``fooling" \cite{fooling} or ``rubbish" \cite{rubbish} data. Therefore, the crux of open-set recognition is to allow rejection of artifacts ``fooling" and estimate the probability of an input being from an unseen class \cite{1vsset}.
	
	Bendale introduced the meta-recognition concept to deep learning, which rejects fooling data by OpenMAX layer \cite{openmax}. And he defined an ``open-world recognition" problem \cite{openworld}, extending nearest class mean type algorithms (NCM) \cite{ncm1,ncm2} to the nearest non-outlier (NNO) algorithm, which not only can identify unseen classes but also learn new classes to achieve incremental learning. Ge Z. introduced a generative adversarial network (GAN) to establish an explicit model and visualize the unseen classes \cite{gan}.	
	
	The end-to-end deep learning methods achieve excellent performance in both RF device identification and open-set of object recognition. However, the open-set RF device identification is still challenging since the signals from different devices have the same structure but slightly different fingerprints. The limitation of deep-learning on open-set RF device identification leads to our effort to apply deep-learning-based classifiers on the open-set RF device identification.

\section{The Proposed Deep Open-Set RF Device Identification (DOS RF-I) Method}\label{method}

\begin{figure*}[htb]
		\centering
		\includegraphics[width=0.9\textwidth]{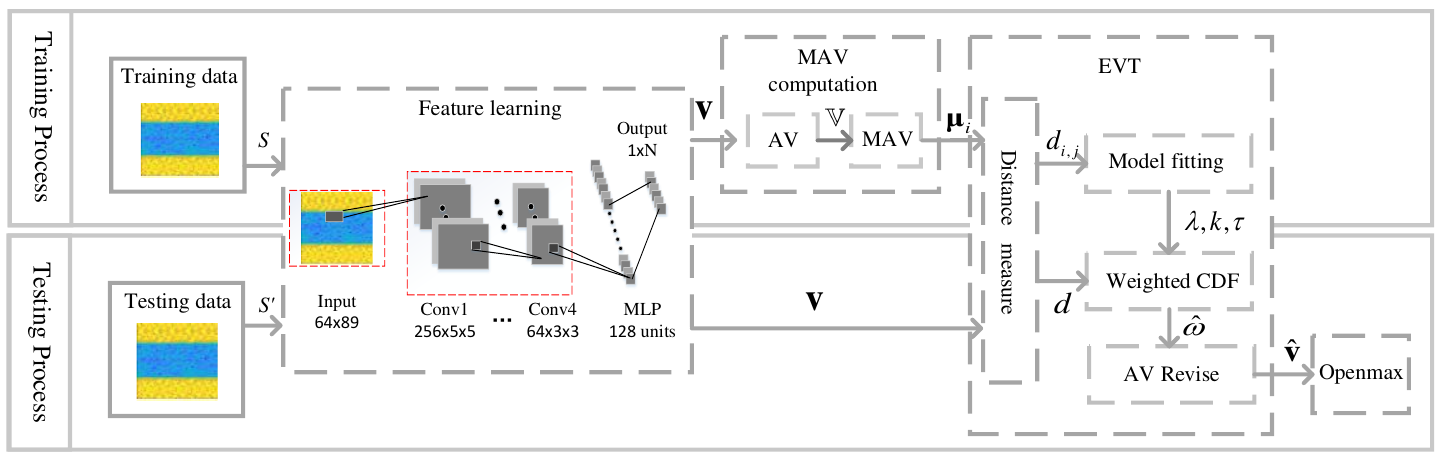}
		\caption{DOS RF-I procedure.}
	\label{overall}
	\end{figure*}

We propose a deep open-set RF device identification (DOS RF-I) method, as shown in Fig. \ref{overall}.
Via extensive experiments for closed-set RF devices identification, as described in section \ref{preprocess}, we selected the STFT pre-processing scheme.
Then, we design deep open-set neural networks to extract class-specific features for known class classification and general features for unseen class identification. The extracted feature is defined as the activation vectors (AV) \cite{openmax}, which carry the feature information of training samples, but some are incorrectly classified. To this end, we collect the AVs of correctly classified training samples and compute the mean activation vectors (MAV) of the AVs to indicate each class's center. Then by modeling the distances of peripheral feature values between the training samples and the MAV, associating with the known class bounds, we have a better expression of known class bounds.
In the testing process, we will calculate the unseen feature values and estimate the open-set probability accordingly.
\subsection{Pre-processing}
Because samples are selected from the output of RF devices with the same type and configuration, there are only slight differences among all outputs. This makes the RF device classification algorithms challenging even under close-set condition. An effective pre-processing scheme enlarges the difference among RF device outputs in deep feature space.



Short-time Fourier transform (SFTF) shows state-of-the-art performance among different pre-processing schemes \cite{wang2019transferred}.
We extract the signal's time-frequency feature via STFT as
 	\begin{equation}
	\centering
	\label{stft}
	\operatorname{STFT}\left( \omega,\xi \right)=\int_{-\infty }^{+\infty }{f\left( t \right)}g\left( t-\xi \right){{e}^{-j \omega t}}dt,
	\end{equation}
where $g(t-\xi)$ is the time window function; $\omega$ is the frequency.

\begin{figure}[htb]
		\centering
		\subfloat[Device 1]{
				\includegraphics[width=0.5 \columnwidth]{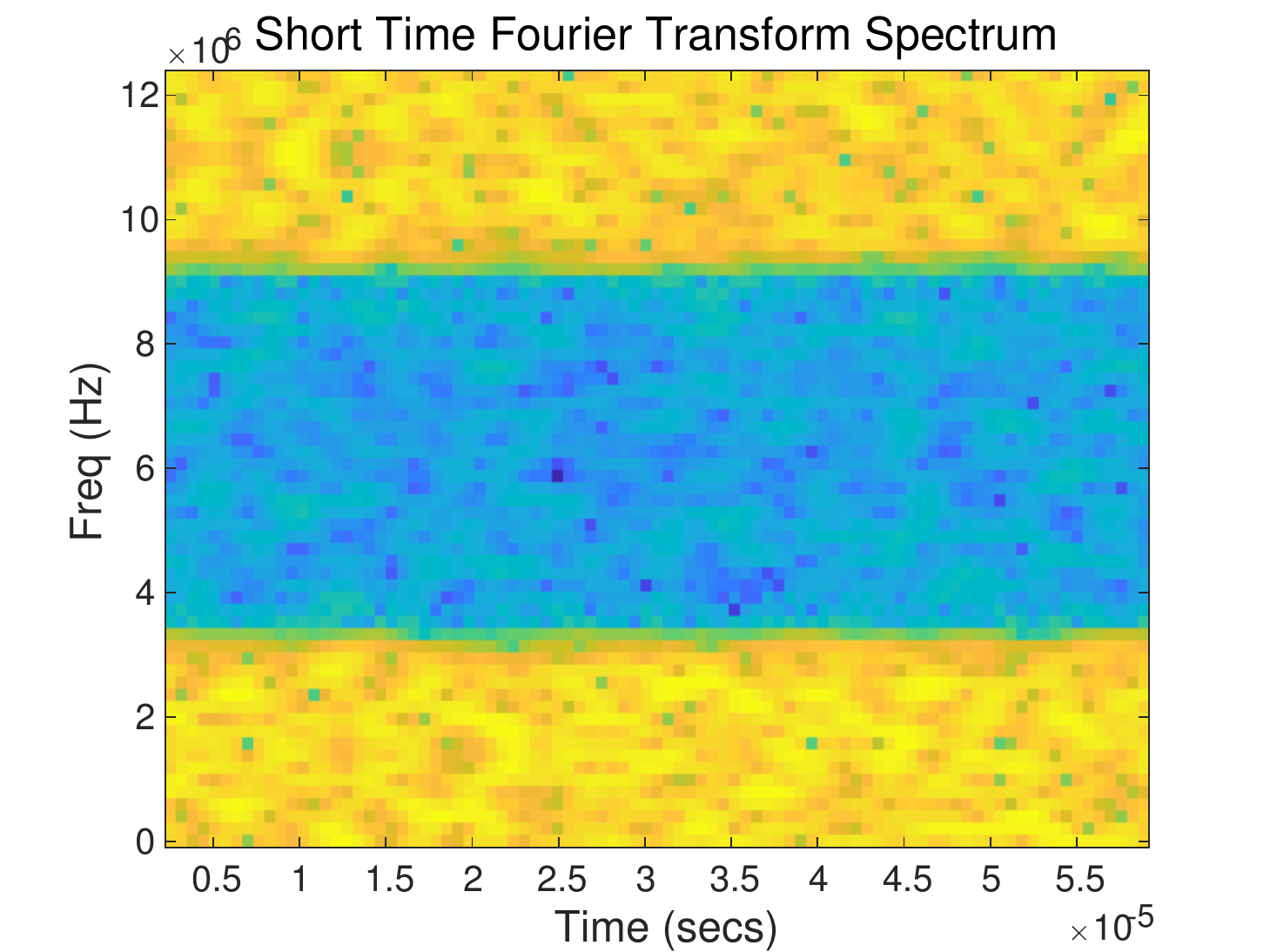}
		}	
		
		\subfloat[Device 2]{
				\includegraphics[width=0.5 \columnwidth]{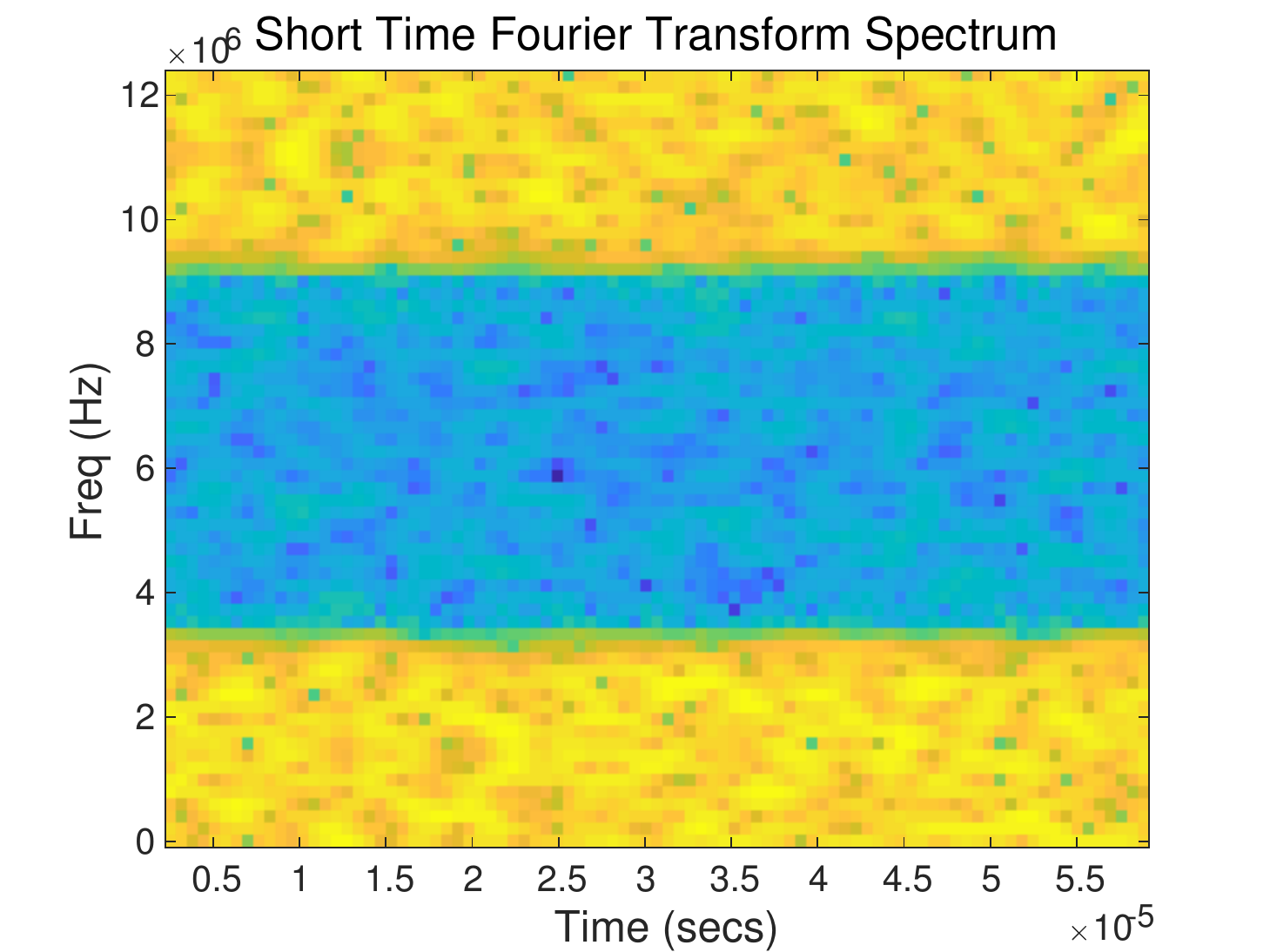}
		}
		\caption{The time-frequency diagram of two typical devices after STFT.}
		\label{stftoutput}
	\end{figure}

To this end, we first recorded and sliced received signal using a $768$-points rectangular window. By applying STFT to each segment, we have the two-dimensional joint distribution diagram with the size of $64\times89$, representing the signal's energy density or intensity at different times and frequencies. As shown in Fig. \ref{stftoutput}, two different devices present different time-frequent diagrams after STFT.
	
After the pre-processing, half of the known class samples are chosen as the training set, while the rest known class samples and all the unseen class samples are treated as the testing set.

\subsection{Feature Extraction}

To extract class-specific features for known class classification and general features for unseen class identification, we proposed a deep open-set neural network, namely open-set CNN, which is based on the deep neural network we proposed in \cite{wang2019transferred}.

\begin{figure}[htb]
	\centering
	\includegraphics[width=0.9\columnwidth]{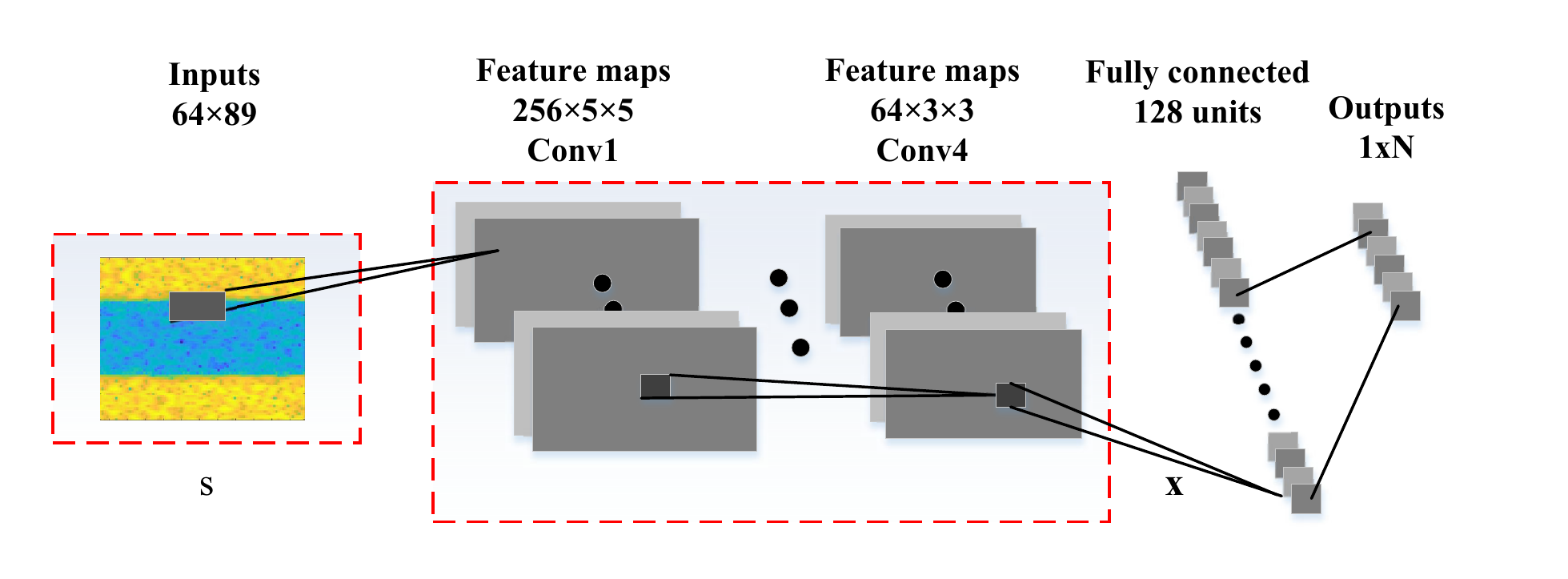}
	\caption{CNN architecture (convolutional kernel size: 5$\times$5, 5$\times$5, 3$\times$3, 3$\times$3).}
	\label{cnn}
	\end{figure}

 The CNN is equipped with four convolutional layers and two fully connected layers, as shown in Fig. \ref{cnn}.
	
The first convolution layer extracts shared low-level features. 
The following three convolution layers sequentially learn features, which convey fingerprint knowledge (i.e., the frequency offset, phase noise, non-linear distortion of the power amplifier, etc). Each convolution layer is bundled with a batch normalization (BN) layer \cite{BN} to accelerate convergence and avoid over-fitting.

The first fully connected layer encodes all features into an $\alpha$-dimensional vector ${\bf{x}}=[{x_1},{x_2},\dots,{x_\alpha}]$ and outputs an $\alpha$-dimensional vector $\widetilde{\bf{x}}=[\widetilde{x_1},\widetilde{x_2},\dots,\widetilde{x_\alpha}]$, whose element is calculated by rectified linear (ReLU) function.
The last fully connected layer with a Softmax operation estimates the probability distribution of belonging to known classes.
The detailed parameters of each layer in CNN are shown in Tab \ref{oscnn}.
	\begin{table}[h]
    \caption{CNN Network parameters.
    \label{oscnn}}
    \centering
    \begin{tabular}{cc}
    \hline
    \textbf{Layer} & \textbf{ Convolution kernel size} \\
    \hline
    Conv1 &    $256\times5\times5$ \\
    Conv2 &    $128\times5\times5$ \\
    Conv3 &    $64\times3\times3$ \\
    Conv4 &    $64\times3\times3$ \\
    \hline
    \end{tabular}
    \end{table}

In the training process, we save the output of the first fully connected layer in Relu as \textit{activation vector} (AV) ${\bf{v}}({\bf{s}})$, whose element ${\bf{v}}_{i,j}$, the $j$-th AVs of $i$-th class, is the feature scores of the penultimate layer defined as
	\begin{equation}
	\centering
	\label{defineav}
	{\bf{v}}_{i,j} = \widetilde{\bf{x}}(s_{i,j}),
	\end{equation}
where ${\bf{s}}=\{s_{i,j}\}$, $(i=1, \dots, \alpha,$ $j = 1, \dots, N^{tr}_{i})$ denotes the collection of training samples; $\alpha$ is the number of classes in the training set; $N^{tr}_{i}$ is the number of samples in each class of training set, that is to say, each sample in training set will output one AV; $\tilde{\bf{x}}(\cdot)$ is the Relu function.
After feature learning, we have AVs of all the training samples.

\subsection{Mean Activation Vectors (MAV) Computing}

AVs carry the feature information of training samples, but some samples are incorrectly classified. We collect the AVs of correctly classified training samples into a new set as
	\begin{equation}
	\begin{split}
	\centering
	\label{rightav}
	\mathbb{V}= \{ {\bf{v}}_{i,j} \mid &\emph{label} \left[s_{i,j} \right] = \mathop{\emph{argmax}}\limits_{i} \left[ {\bf{v}}_{i,j} \right], \\
	&i=1,\dots, \alpha, j=1,\dots, r_i \},
	\end{split}
	\end{equation}
where the operation $\emph{label}[\cdot]$ denotes getting the label of a specific sample; $r_i$ denotes the number of correctly classified training samples in $i$-th class.

To indicate the center of each class, we define the \textit{mean activation vectors} (MAVs) as
	\begin{equation}
	\centering
	\label{mav}
	{\bm{\mu}}_i= \frac{1}{r_i}\sum_{j=1}^{r_i} {\bf{v}}_{i,j}, {\bf{v}}_{i,j} \in \mathbb{V}.
	\end{equation}

\subsection{Distance Measurement}
We then measure the distance $d_{i,j}$ between the AVs and the corresponding MAV, defined as the combination of weighted Euclidean distance and cosine distance, where Euclidean distance measures the direct distance between two features and Cosine distance measures the difference in direction between two features. The combination of these two distance measures reflect the relationship between features, given by
	\begin{equation}		
	\label{distance}
	\centering
	d_{i,j}= \beta{\parallel {\bf{v}}_{i,j} - {\bm{\mu}}_i \parallel_{2}}
	-\frac{{\bm{\mu}_i} \cdot {\bf{v}}_{i,j}}{{\parallel{\bm{\mu}}_i \parallel_{2}}{\parallel {\bf{v}}_{i,j}\parallel_2}}+1, {\bf{v}}_{i,j} \in \mathbb{V},
	\end{equation}
where $\beta$ is the weighting factor.

In our open-set RF device identification problem, the ranges of AVs in different classes are overlapped. Still, the probability density distribution of AVs in different classes are quite different. Thus we model the Weibull distribution of AVs as a better expression of known class bounds, making the unseen class identification more accurate.

\subsection{The Extreme Value Theory (EVT) based Peripheral Points Management and Weibull Distribution Modeling}

Often, the open-set recognition system based on hard threshold method cannot accurately discriminate the sample points near the decision boundary, and thresholds need to be set artificially, which inevitably leads to errors. Fortunately, Extreme Value Theorem (EVT) \cite{evt} can estimate detection thresholds and remove the influence of outliers in the sample simultaneously \cite{5229148}. To overcome the above shortcomings, in this paper, we introduce the statistical EVT to decide the soft threshold and manage the peripheral points, which refer to the points far away from the MAVs.
%

%
%

Let $(d_{i,1},d_{i,2},\dots)$ in \eqref{distance} be a sequence of the distance, which is i.i.d., and $M_{r_i}=max\{d_{i,1},d_{i,2},\dots,d_{i,r_i}\}$. If a sequence of pairs of real numbers $(a_{r_i},b_{r_i})$ exists such that each $a_{r_i}>0$ and
 \begin{equation}
 \centering
 \label{evteq}
 \lim_{d \to \infty}P(\frac{M_{r_i}-b_{r_i}}{a_{r_i}}\leq d)=F(d),
 \end{equation}
then if $F(d)$ is a non-degenerate distribution function, it belongs to one of three extreme value distributions, i.e., Gumbel (I), Frechet (II), and Reversed Weibull (III) distributions. Gumbel and Frechet work for unbounded distributions, and Weibull works for bounded. For most recognition systems, the distance or similarity scores are bounded with both upper and lower. Besides, previous works \cite{yuan,weibull} have shown open-set sample distributions follow Weibull distribution.
Therefore, we use the Weibull model to fit the distribution of the extreme values, i.e., rare events, and quantitatively provide the probability of a certain distance in each class.
The Weibull probability density function (PDF) is defined as
	\begin{equation}
	\centering
	\label{weibullmodel}
	\begin{split}
	\omega\left( d;\lambda,k,\tau \right) =
	\begin{cases}
	\frac{k}{\lambda} \left( \frac{d-\tau}{\lambda}\right)^{k-1} e^{-\left( \frac{d-\tau}{\lambda}\right)^k}	&d\ge 0, \\
	0	&d<0,
	\end{cases}
	\end{split}
	\end{equation}
where $k$ is the shape parameter; $\lambda>0$ is the scale parameter for PDF curve magnifying or shrinking; $\tau$ is the location parameter.
Accordingly, the Weibull CDF is defined as
	\begin{equation}
	\centering
	\label{weibullcdf}
	\begin{split}
	\Omega\left( d;\lambda,k,\tau \right) =
	\begin{cases}
	1 - e^{-\left( \frac{d-\tau}{\lambda}\right)^k}	&d\ge 0, \\
	0	&d<0.
	\end{cases}
	\end{split}
	\end{equation}

As shown in Fig. \ref{overall}, we use the Weibull models to train the deep network to extract the inter-class disparity features for known classes classification and the intra-class shared features for unseen class identification. By modeling the tail of the distance distribution between AVs and the corresponding MAV, we have the per class Weibull model $\omega_i\left( d;\lambda,k,\tau \right)$ and $\Omega_i\left( d;\lambda,k,\tau \right)$, $i=1,2\dots\alpha$.

\begin{figure}[htb]
		\centering
		\includegraphics[width=3in]{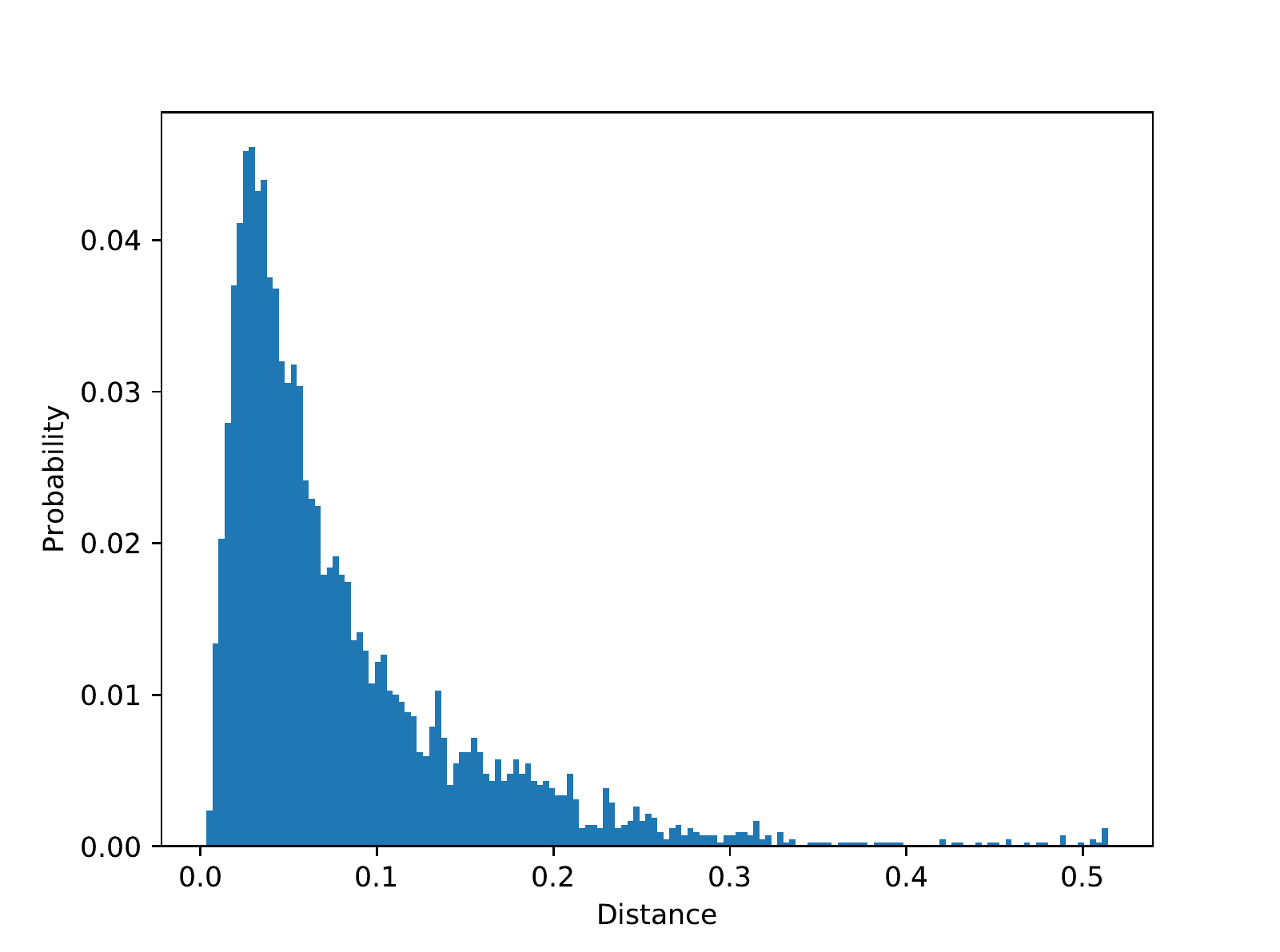}
		\caption{Distance distribution of the 01 device (\textbf{V-set}).}
		\label{01distance}
	\end{figure}

\begin{figure}[htb]
	\centering
	\includegraphics[width=3in]{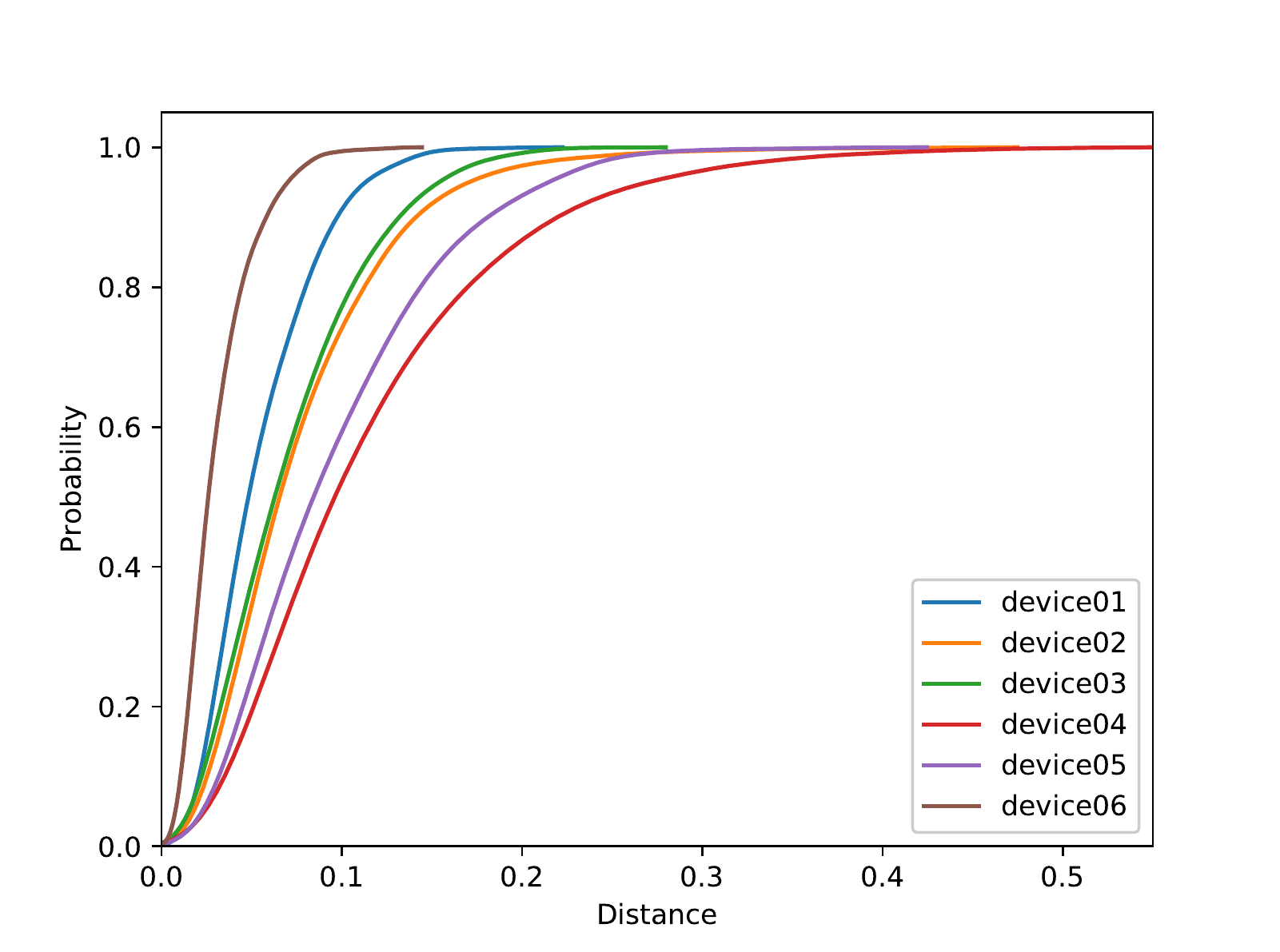}
	\caption{CDF curves of 6 devices (\textbf{V-set}).}
	\label{cdf}
	\end{figure}

Taking the distance distribution of the $01$ device as an example, Fig. \ref{01distance} shows the distribution histogram of all AVs in the $1^{st}$ class, where the distance equals to $0$ representing the MAV of the $01$ device. In practical terms, to highlight the peripheral points, we use the LibMR \cite{yuan} FitHigh function to fit the $20$ tail samples in the $1^{st}$ class, obtaining the Weibull PDF and CDF of the $01$ device, i.e., $\omega_1$ and $\Omega_1$, respectively. Following the same steps, we have $\omega_i$ and $\Omega_i$ of each class, as shown in Fig. \ref{cdf}.

As a result, in the testing process, the \textit{open-set AV} can be calculated accordingly using \textit{Weighted CDF} ${\Omega}^{\prime}_i$, which will be illustrated in the later subsection.

\subsection{Open-Set AV Calculation}
During training and testing, half samples of the known classes are chosen as the training set $\bf{s}$, while the rest known classes samples and all the unknown classes samples are treated as the testing set $\bf{s}^{\prime}$. In the testing process, we first measure the distance vector of testing samples $\bf{s}^{\prime}$ according to \eqref{defineav}-\eqref{distance}, i.e., ${\bf{d}}=[d_1,d_2,\dots,d_\alpha]$, where $d_i$ is the distance between the AVs of testing samples $\bf{v}(s^{\prime})$ and the MAV of $i$-th class $\bm{\mu}_i$ in \eqref{mav}.

Then we define the revised \textit{Weighted CDF}, ${{\bm{\Omega}}^{\prime}(\bf{d})}=[{\Omega}^{\prime}_1,{\Omega}^{\prime}_2,\dots,{\Omega}^{\prime}_i]$, where the element ${\Omega}^{\prime}_i\left( d,\eta_i;\lambda,k,\tau \right)$ is defined as
	\begin{equation}
	\centering
	\label{reviseparameter}
	\begin{split}
	{\Omega}^{\prime}_i\left( d,\eta_i;\lambda,k,\tau \right) =
	\begin{cases}
	1 - \frac{\alpha-\eta_i}{\alpha}e^{-\left( \frac{d-\tau}{\lambda}\right)^k}	&d\ge 0, \\
	0	&d<0;
	\end{cases}
	\end{split}
	\end{equation}
$\eta_i$ is the weighting parameter to scale the Weibull CDF given by
	\begin{equation}
	\centering
	\label{weighting}
	   \eta_i= argsort[v_i({\bf{s}}^{\prime})];
	\end{equation}
where the operation $argsort$ returns the indices that sort the vector $v_i({\bf{s}}^{\prime})$ in order.

Comparing with AV, we add the $0$-th dimension $\hat{v}_0(s^{\prime})$ to indicate the unseen class feature. Thus, we have the \textit{open-set AV} $\hat{\bf{v}}({\bf{s}}^{\prime})$, whose element $\hat{v}_{i^{\prime}}({\bf{s}}^{\prime})$ is defined as
\begin{equation}
	\centering
	\label{opensetav}
	\begin{split}	
	\hat{v}_{i^{\prime}}({\bf{s}}^{\prime}) =
	\begin{cases}
	v_{i^{\prime}}({\bf{s}}^{\prime}) \circ  {\Omega}^{\prime}_{i^{\prime}} &i^{\prime}=1,\dots\alpha, \\
	\sum_{{i}=1}^{\alpha} v_{i}  \left( 1-{\Omega}^{\prime}_{i}  \right) &{i^{'}}=0,
	\end{cases}
	\end{split}
	\end{equation}	
where $i^{'}=1,\dots,\alpha$ is the index of the open-set class; $i^{'}=0$ is the index of the unseen class; $\circ$ is the Hadamard product.
	
\subsection{Open-Set Identification using Openmax Function}
After obtaining the open-set AVs, i.e., $\hat{\bf{v}}({\bf{s}}^{\prime})$, we can estimate the open-set probability of samples using the Openmax function, defined as
	\begin{equation}
	\centering
	\label{openmax}
	\begin{split}
	\hat{P}\left( y=i^{\prime}|\hat{\bf{v}} \right) = \frac{e^{\hat{v}_{i^{\prime}}}}{\sum\limits_{i^{\prime}=0}^\alpha{e^{\hat{v}_{i^{\prime}}}}}.
	\end{split}
	\end{equation}

In summary, by analyzing the post-recognition scores using EVT, we have the open-set probability according to \eqref{openmax}, which is used to identify the unseen IoT devices. In Fig. \ref{overall}, we describe the algorithm flow in detail.

\section{Experiment}\label{experiment}
We conduct the experiments and numerical analysis to evaluate our proposed DOS RF-I method. The setup is described as follows. We build our model based on the Tensorflow platform \cite{tensorflow} and Keras \cite{keras}. The adaptive moment estimation (Adam) optimizer \cite{Adam} is applied over the training set with a batch size of 128 and the max epoch of $100$. The model is trained on a workstation with an Intel $I7-8700$ CPU and an NVIDIA Geforce GTX $1080$ GPU.

\subsection{Dataset Construction}
We use six RF devices with the same type as the terminal nodes in the perception layer. Each RF device transmits the same kinds of signals. We construct two datasets by sampling a voice signal set (\textbf{V-Set}) and a data signal set (\textbf{D-Set}). 
The \textbf{V-Set} is composed of voice signals transmitted by six RF devices with a similar type and frequency. Each sample is a record of the signal received by the RF devices when transmitting the same audio clip. The intervals in the samples represent voice pauses and altering amplitude, indicating the changes in volume. The \textbf{D-Set} contains the data transmitted through the communication link, which does not have apparent gaps. Since all data is selected from six devices, the two datasets contain six classes, consisting of 7000 samples. Each sample includes I/Q components. We randomly select $\alpha$ classes as the known classes, which take part in the training process. The rest $6-\alpha$ classes were set as unknown/unseen classes.

\subsection{Preliminaries}
Generally, for open-set conditions, identifying unseen classes in the testing set is a big challenge due to the incomplete knowledge in the training process.
More specifically, in our experiments, the open degrees represent the ratio of unseen classes. Therefore, evaluating overall accuracy or open recognition accuracy will receive different results since the accuracy is influenced by the open degree. We select multiple stable metrics for a fair comparison to provide a valid evaluation for the proposed method.
\subsubsection{Openness}
Let us define the ``openness'' \cite{1vsset} to measure the open degree as
	\begin{equation}
	\centering
	\label{openness}
	\begin{split}
	\emph{\mbox{Openness}}=1-\sqrt{\cfrac{2\times N_{TA}}{N_{TG}+N_{TE}}},
	\end{split}
	\end{equation}
	where $N_{TA}$ is the number of training classes (known classes); $N_{TE}$ and $N_{TG}$ are the numbers of the testing classes and the target classes, which contain known and unknown classes, and known and unknown classes that differ from the testing class, respectively. $\emph{\mbox{Openness}}=0$ refers to a closed-set problem, i.e., the number of classes in the testing set is the same as in the training set, and the target classes is the same as the testing classes. The larger the value of ``openness'', the more open the problem.
	
\subsubsection{Accuracy}
The recognition accuracy includes closed recognition accuracy (C-accuracy), open recognition accuracy (O-accuracy), and overall recognition accuracy (A-accuracy). C-accuracy (O-accuracy) denotes the correctly classified known (unseen) samples among all known (unseen) samples, defined as
	\begin{equation}
	\centering
	\label{C-accuracy}
	\begin{split}
	&\emph{\mbox{C-accuracy}}=	\cfrac{\emph{\mbox{KP}}}{\emph{\mbox{KP+KN}}},\\
	&\emph{\mbox{O-accuracy}}=	\cfrac{\emph{\mbox{NP}}}{\emph{\mbox{NP+UN}}}.
	\end{split}
	\end{equation}
where KP denotes the correctly classified known samples; KN is the incorrectly classified known samples; UP/UN are the correctly/incorrectly classified unseen samples.
	
A-accuracy is defined to balance the C-accuracy and O-accuracy, indicating the overall system performance, i.e., correctly classified samples among all samples, given by
	\begin{equation}
	\centering
	\label{A-accuracy}
	\begin{split}
	&\emph{\mbox{A-accuracy}}=\cfrac{\emph{\mbox{KP + KN}}}{\emph{\mbox{KP + KN + UP + UN}}}.
	\end{split}
	\end{equation}

\subsection{Impact of Pre-processing Scheme}\label{preprocess}
To show the effectiveness of our pre-processing module, we compare the closed-set classification performance using different pre-processing schemes (shown in Tab \ref{preprocessing}). The STFT pre-processing module improves the classification performance by around $15.1 \%$ comparing with the base model (original IQ samples with only CNN). We also adopt our pre-processing module to other neural network architectures, i.e., Two-Channel Neural Network (TCNN) \cite{wang2019transferred}.
The results also demonstrate that the STFT pre-processing is effective for both CNN and TCNN.
    \begin{table}[h]

    \caption{Closed-set classification performance.}
    \label{preprocessing}
    \centering
    \begin{tabular}{cc}
    \hline
    \textbf{Statistical parameters} & \textbf{Classification accuracy} \\
    \hline
    Original IQ samples (CNN) &     $79.8\%$ \\
    STFT (CNN) &  $94.9\%$ \\
    0-order FD (CNN) &      $85\%$ \\
    5-order FD (CNN) &      $86\%$ \\
    9-order FD (CNN) &      $86\%$ \\
    3-order cumulant (CNN) &      $55\%$ \\
    \hline
    Orignal IQ samples (TCNN) &  $82.1\%$ \\
    STFT (TCNN) &  $96.2\%$ \\
    \hline
    \end{tabular}
    \end{table}

\subsection{Comparison between Openset Techniques}
Recall the threshold on the distance (hard threshold discriminant) defined in \eqref{distance} that determines where a sample is an unseen class, and we compare the hard threshold discriminant with the proposed DOS RF-I method.

\subsubsection{Openness}
The performances of the hard threshold discriminant and the proposed DOS RF-I method are summarized in Tab \ref{hardandour}. The proposed DOS RF-I method is far superior to the hard threshold discriminant. With the increment of openness, the performance gap is gradually enlarged (from $12.97\%$ to $29.25\%$).

\renewcommand\arraystretch{1.5}
\begin{table}[h]
	\caption{A-accuracy on \textbf{V-set}.
	\label{hardandour}}
	\begin{center}
		\begin{tabular}{cccc}
			\hline
			Method & 1 Unseen & 2 Unseen & 3 Unseen \\
			\hline
			Hard Threshold    & 81.27\% &  79.99\% &  64.18\% \\
			\hline
			DOS RF-I method 	& 94.24\% & 94.85\% & 93.43\% \\
			\hline
		\end{tabular}
	\end{center}
\end{table}

\subsubsection{Threshold}
\begin{figure}[htb]
 		\centering
 		\includegraphics[width=3in]{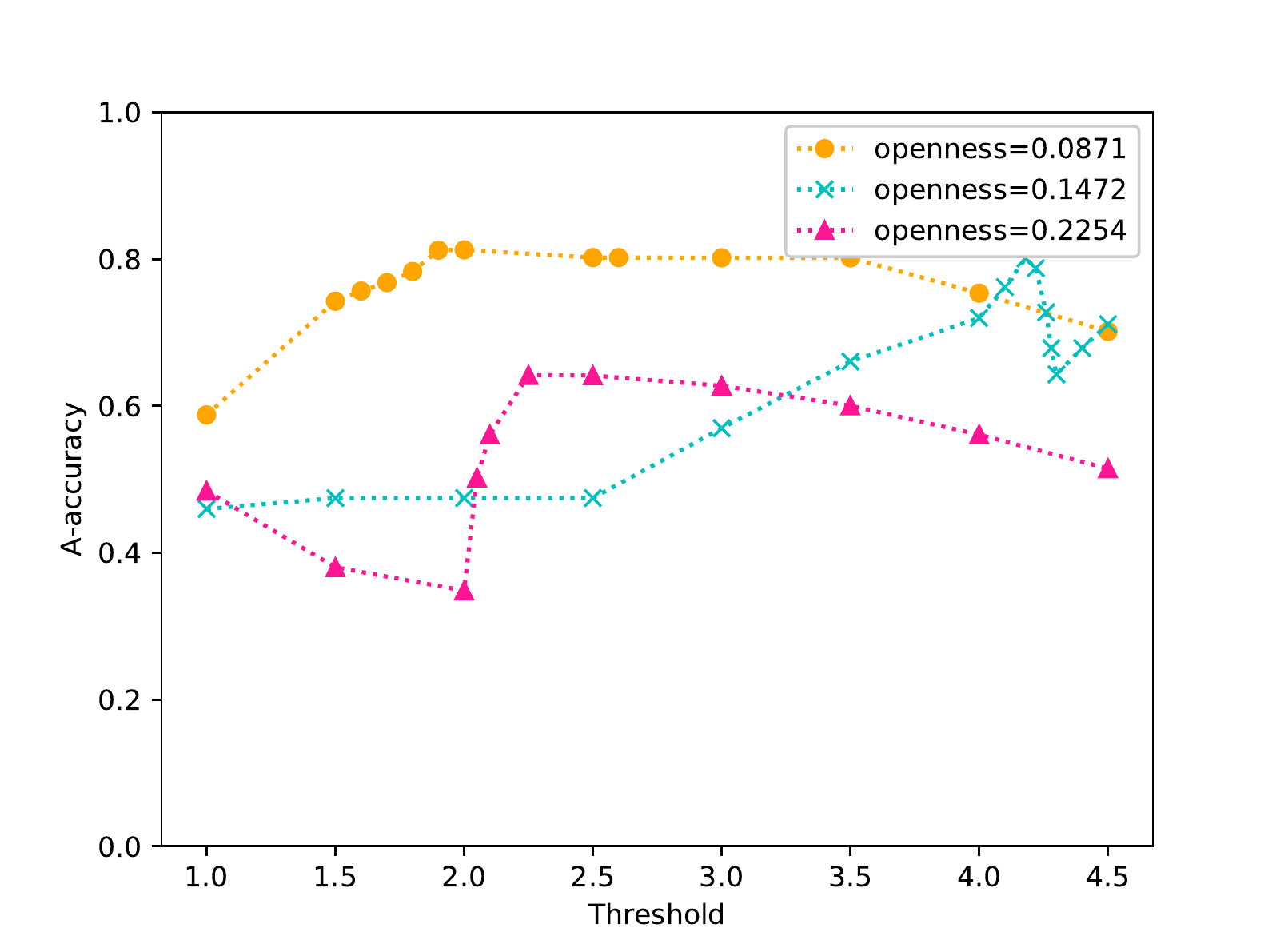}
 		\caption{A-accuracy with different thresholds for \textbf{V-set} identification. (hard threshold discriminance).}
 		\label{all_threshold}
 		\end{figure}

The DOS RF-I uses the Weibull distribution of AVs to fit the known class bounds instead of setting a hard threshold. Therefore, DOS RF-I will not be affected by the threshold and present the known class bounds more tightly. On the contrary, the hard threshold discriminant is sensitive to the threshold. Noted that the best performance of hard threshold discriminant is measured under A-accuracy with altering thresholds ($1.0\sim4.5$). As shown in Fig. \ref{all_threshold}, the optimum threshold for the maximum A-accuracy varies with different openness settings.

With the threshold increasing, fewer samples are classified as unseen classes, while more samples are identified as known classes. Therefore, it is not feasible to determine a hard and unadjustable threshold to achieve the best classification performance in practical applications.

\subsection{Impact of Dataset}
The proposed DOS RF-I method has an robust performance on both \textbf{V-Set} and \textbf{D-Set}. As shown in Tab \ref{vdopensetacc}, the O-accuracy and A-accuracy of \textbf{D-Set} are slightly higher than \textbf{V-Set} by around $5\%$. In addition, the C-accuracy of \textbf{V-Set} and \textbf{D-Set} using Softmax are $94.87\%$ and $99.94\%$, respectively. The performance on voice signals is lower because even if the content of the speech is the same, there is no guarantee that the signals are identical.

\renewcommand\arraystretch{1.5}
\begin{table}[h]
	\caption{Recognition performance of the \textbf{V-set} and the \textbf{D-set} using the DOS RF-I method.}
	\label{vdopensetacc}
	\begin{center}
		\begin{tabular}{|cc|cc|cc|}
			\hline
			\multicolumn{2}{|c|}{\diagbox{\makecell[c]{Class\\Number}}{DataSet}}&
			\multicolumn{2}{c|}{\textbf{V-set}}&
			\multicolumn{2}{c|}{\textbf{D-set}} \\
			\cline{1-2}\cline{3-4}\cline{5-6}
			Known & Unseen & O-acc & A-acc  & O-acc & A-acc \\
			\hline
			6 & 0 & 94.87\% & 94.87\%  & 99.94\% & 99.94\% \\
			\hline
			5 & 1 & 93.02\% & 94.24\%  & 99.99\% & 99.78\% \\
			\hline
			4 & 2 & 98.57\% & 94.85\%  & 97.65\% & 99.07\% \\
			\hline
			3 & 3 & 93.35\% & 93.43\%  & 99.99\% & 99.89\% \\
			\hline
		\end{tabular}
	\end{center}
\end{table}

\subsection{Impact of Network Structure}
\begin{figure}[h]
	\centering
	\includegraphics[width=3in]{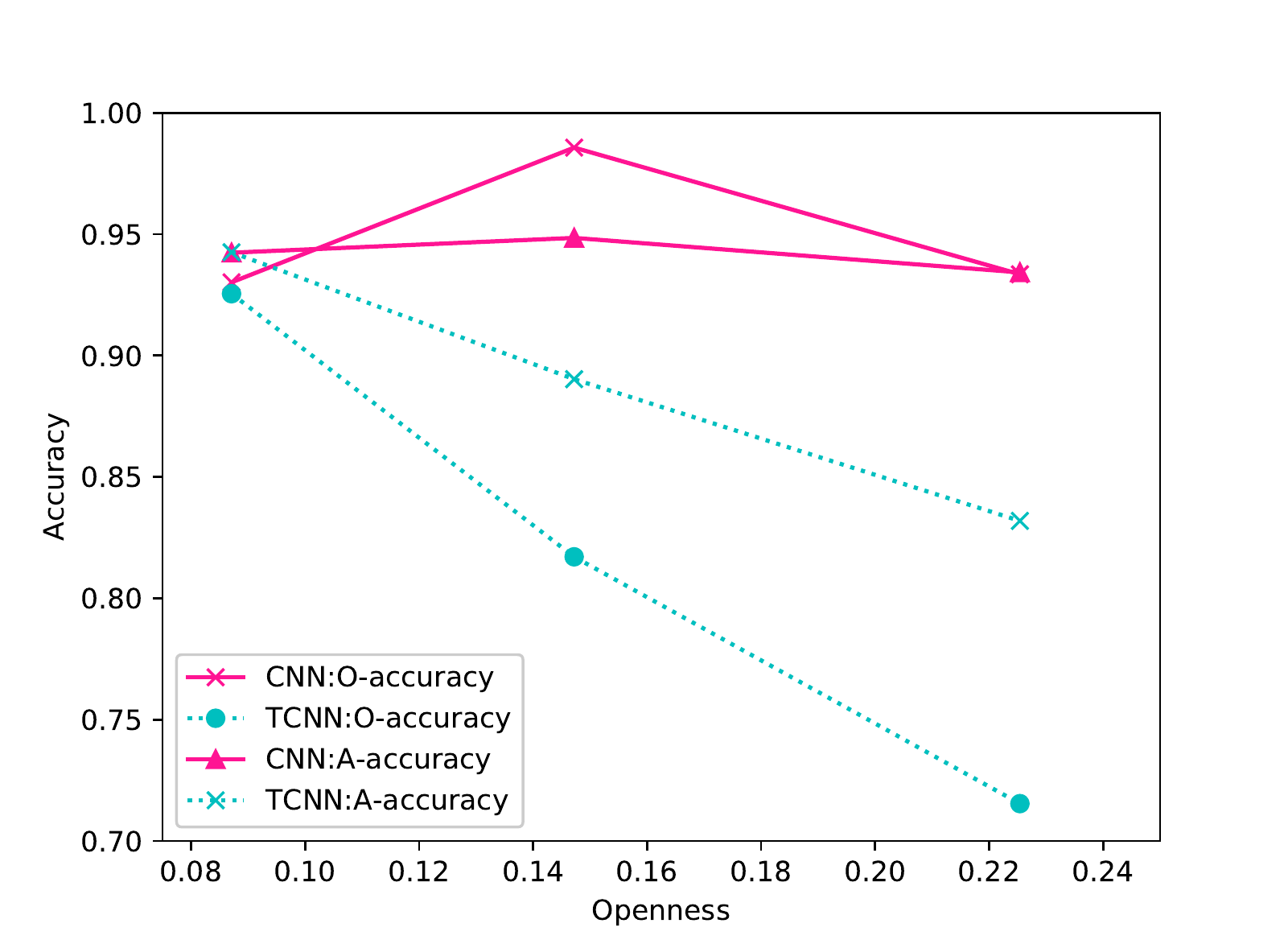}
	\caption{\textbf{V-set} recognition accuracy of different neural network structures. }
	\label{cnntcnn}
	\end{figure}

To evaluate the contribution of different neural network structures (CNN and TCNN \cite{wang2019transferred}), an ablation study is conducted on the \textbf{V-Set}.
The results based on recognition accuracy are shown in Fig. \ref{cnntcnn}.
The O-accuracy of TCNN is much less than CNN. With the increase of openness, O-accuracy significantly decreases by $0.47\% \sim 21.81\%$. Due to the sharp decline of O-accuracy, the A-accuracy of TCNN also dropped to $10.24\%$.

TCNN with a complex structure can extract rich, in-depth features. Still, the open-set identification requires the algorithm to focus on low-level features and be insensitive to outlier features from unseen classes.

\subsection{Impact of Distance Metric}
We analyze the impact of different distance metrics in this subsection. The recognition performance of $3$ unseen classes in the \textbf{V-set} using CNN with different metrics is summarized in Tab \ref{distancemeasure}.
	
In Tab \ref{distancemeasure}, the method combined Euclidean and Cosine distance achieves the best performance on both O-accuracy ($93.35\%$) and A-accuracy ($93.34\%$) as it exploits the similarity. In contrast, the distance metrics such as cosine or Canberra are good at measuring geometrical distance but cannot represent the similarity of different AVs, resulting in poor performance in open-set identification problems.

\renewcommand\arraystretch{1.5}
\begin{table}
	\caption{Recognition performance of 3 unseen classes \textbf{V-set} using CNN with various distance metrics.}
	\label{distancemeasure}
	\begin{center}
		\begin{tabular}{cccc}
			\toprule[1pt]
			Distance metrics & C-accuracy & O-accuracy & A-accuracy \\
			\midrule[1pt]
			Euclidean & 93.56\% &  89.21\% &  91.38\% \\
			\hline
			Cosine & 93.44\% &  44.36\% &  68.90\% \\
			\hline
			Euclidean + Cosine  & 93.52\% &  \textbf{93.35\%} &  \textbf{93.43\%} \\
			\hline
			Canberra & 93.42\% &  66.74\% &  80.08\% \\
			\hline
			Chebyshev & \textbf{93.58\%} &  89.90\% &  91.74\% \\
			\hline
			Minkowski & 93.56\% &  89.21\% &  91.38\% \\
			\hline
		\end{tabular}
	\end{center}
\end{table}

\subsection{Impact of Tail Size}
\begin{figure}[htb]
	\centering
	\includegraphics[width=3in]{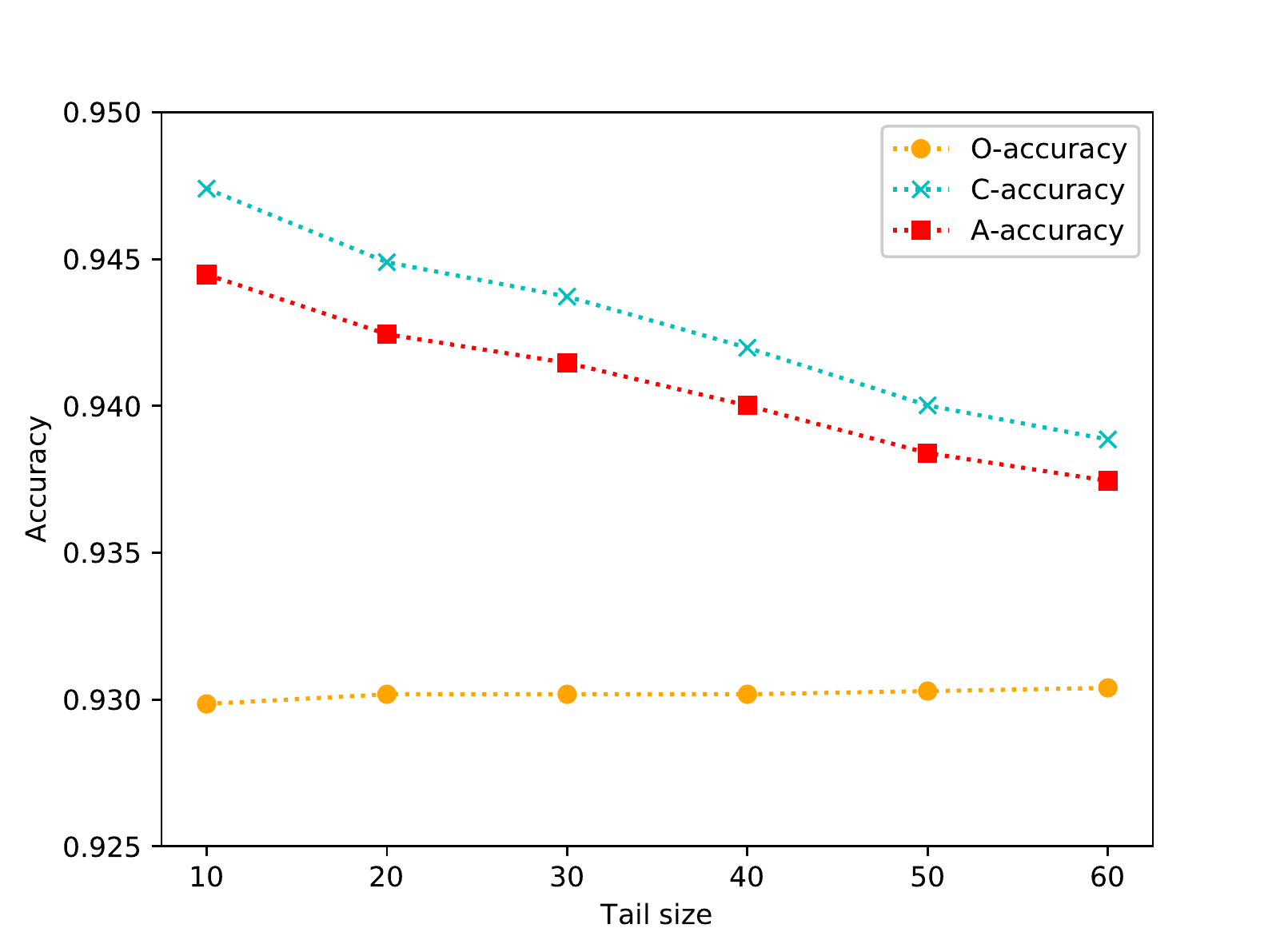}
	\caption{\textbf{V-set} single unseen class recognition accuracy under different tail sizes. }
	\label{1opentailsize}
	\end{figure}

We also study the influence of tail size in Weibull fitting. Fig. \ref{1opentailsize} shows how tail sizes affect the recognition performance of single unseen class in the \textbf{V-set}. As the tail size increases from $10$ to $60$ the O-accuracy remains invariable, but C-accuracy declines by $0.85\%$ (from $94.74\%$ to $93.89\%$), which indicates some known class samples are incorrectly identified as unseen classes.

More boundary points participate in the open-set classification with the larger tail size, and the known classes' boundary moves inward. Because of boundary points' low appearing probabilities, they are more likely to be identified as unseen class samples, which affected the discrimination accuracy of the boundary points in the close-set. 

\subsection{Openness Degree Analysis}
To measure the openness degree and its influence on O-accuracy, we use a broken line graph to display the recognition performance of the \textbf{V-Set} and the \textbf{D-Set} using the proposed DOS RF-I method with CNN.
			\begin{figure}[htb]
		\centering
		\includegraphics[width=3in]{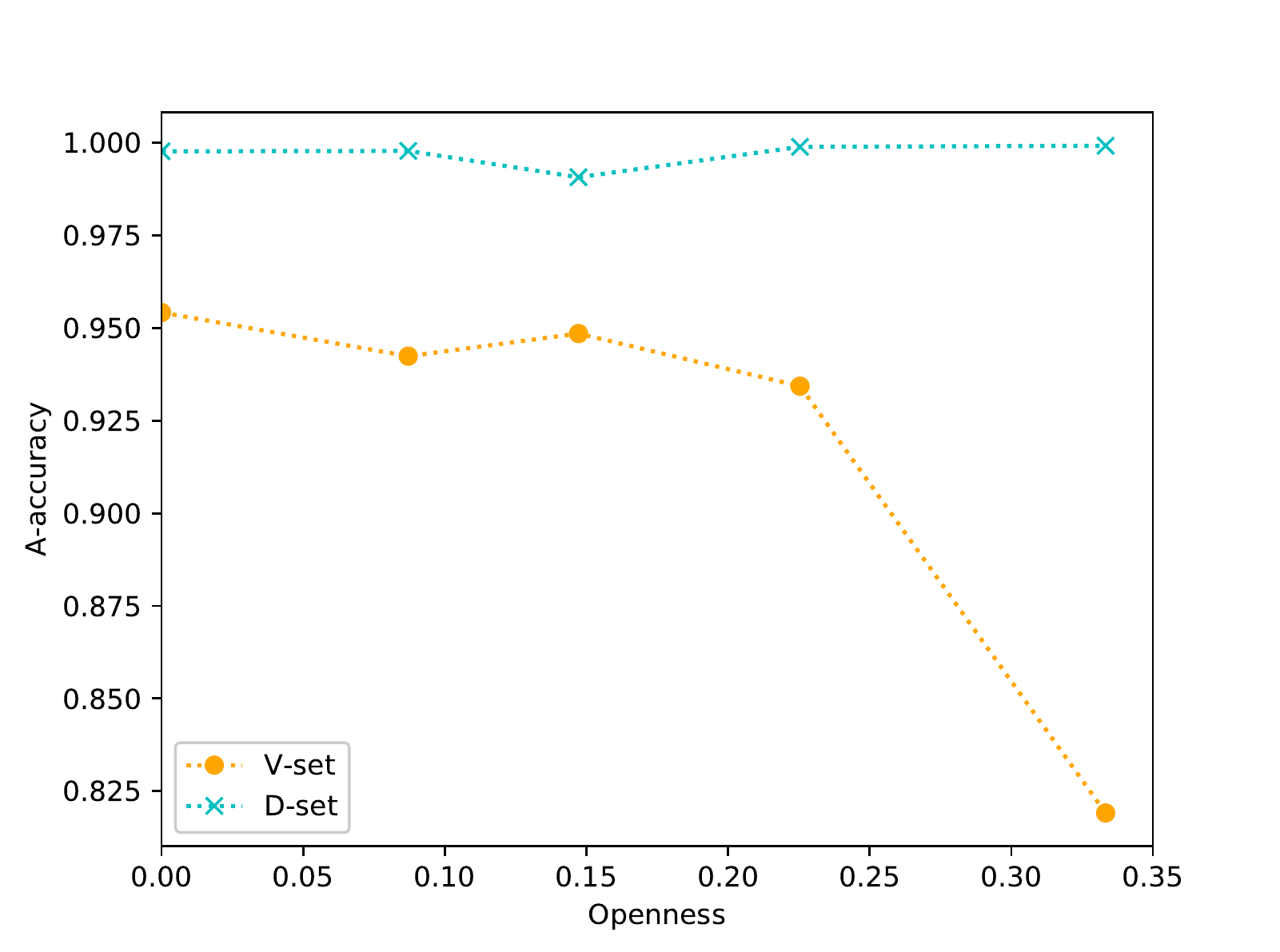}
		\caption{A-accuracy under different openness degrees.}
		\label{opennessvset}
		\end{figure}
As shown in Fig. \ref{opennessvset}, with the increment of openness, the \textbf{V-Set} A-accuracy has a downtrend. Especially when the number of the unseen classes is more than the trained classes (Openness $= 0.33$), the A-accuracy rapidly fell to $81.91\%$.
	
The \textbf{D-Set} recognition accuracy is stable, and the A-accuracy declines by $0.70\%$ when the openness equals to $0.15$. Moreover, with the growth of openness, the A-accuracy is slightly increased. When openness equals to $0.33$ (maximum), the A-accuracy is $99.92\% $, which is $0.15\%$ higher than that under the closed-set condition. Although we do not provide O-accuracy and C-accuracy in Fig. \ref{opennessvset}, we want to point out that the stability trend of A-accuracy for \textbf{D-Set} is coincident with O-accuracy and C-accuracy. This result shows that our system is robust to openness and performs well in both open-set and closed-set identification.

\section{Conclusion}\label{conclusion}
In this paper, a novel meta-learning-based DOS RF-I method is proposed to solve the security problem faced by massive RF secure transmission. An STFT pre-processing module is designed to exploit the slight difference among features learned from the output of various RF devices, and the neural networks with the STFT module achieve greater performance. Moreover, the extreme value theory, i.e., Weibull distribution, was introduced to present the known class bounds more precisely, making the system more robust. The experimental results and comparison with the hard threshold discriminant demonstrate that the proposed DOS RF-I method provide better security and not be affected by the threshold setting.


\end{document}